# Sublimable complexes with spin switching: Chemical design, processing as thin films and integration in graphene-based devices

Miguel Gavara-Edo (ORCID: 0000-0002-0311-3874),[a,d] Francisco Javier Valverde-Muñoz (ORCID: 0000-0003-3578-5445),[a,d] Rosa Córdoba (ORCID: 0000-0002-6180-8113),[a] M. Carmen Muñoz (ORCID: 0000-0003-2630-3897),[b] Javier Herrero-Martín (ORCID: 0000-0002-7015-1635),[c] José Antonio Real (ORCID: 0000-0002-2302-561X)[a]* and Eugenio Coronado (ORCID: 0000-0002-1848-8791)[a]*

[a]Institute of Molecular Science, University of Valencia, Catedrático José Beltrán 2, Paterna 46980, Spain
[b]Departamento de Fisica Aplicada, Universitat Politècnica de València, Camino de Vera s/n, Valencia 46022, Spain
[c]Boreas Beamline CELLS-ALBA Synchrotron, Carrer de la Llum 2-26, Cerdanyola del Vallès 08290, Spain
[d]These authors contributed equally to this work.

ABSTRACT

Among the different types of switchable molecular compounds, sublimable Fe(II) SCO molecules provide a suitable platform to develop smart devices that respond to external stimuli. Here we report the synthesis, crystallographic structure and magnetic properties of three new neutral Fe(II) SCO molecules belonging to the {Fe[H$_2$B(pz)$_2$]$_2$(L)} family with bidentate-$\alpha$-diimine ligands L = 3-(pyridin-2-yl)-[1,2,3]triazolo[1,5-$a$]pyridine (tzpy), 5,5',6,6'-tetrahydro-4$H$,4'$H$-2,2'-bi(1,3-thiazine) (btz) and 4,4',5,5'-tetrahydro-2,2'-bithiazole (bt) (**1**, **2** and **3**, respectively), as well as two solvated forms of **1** and **3**. All three desolvated compounds present thermal- and light-induced SCO transitions with different degrees of cooperativity and effectiveness. Furthermore, **1** and **2** are demonstrated to be sublimable under HV conditions affording homogeneous thin films 200 nm thick (**TF1** and **TF2**) that retain the chemical integrity of the original molecules regardless the deposition surface. The SCO behaviour of the films is characterized by XAS technique revealing the partial retainment of both thermal- and light-induced spin transitions, yet losing the cooperativity. Finally, SCO/2D horizontal hybrid devices based on CVD-graphene are produced using these films. Being the first ones of this type utilizing molecules of {Fe[H$_2$B(pz)$_2$]$_2$(L)} family, with L = tzpy and btz, the devices have allowed the successful detection of the thermal SCO transition through the electric properties of the CVD-graphene.

INTRODUCTION

Octahedral Fe$^{II}$ spin crossover (SCO) complexes are molecular materials that reversibly switch between the high-spin (HS: t$_{2g}^4$e$_g^2$, S = 2) and low-spin (LS: t$_{2g}^6$e$_g^0$, S = 0) states inducing changes in the magnetic, optical, electrical and structural properties controlled through the action of external stimuli such as temperature and/or pressure changes, light irradiation and guest molecules. These features have attracted much attention due to their potential applicability as active components in electronic and spintronic devices, e.g. sensors and memories.[1–9] To this end, selected $Fe^{II}$ SCO materials have been processed at sub-micrometer scales as nanocrystals/nanoparticles or also as thin films usually deposited by layer-by-layer or high-vacuum (HV) sublimation methods.[10,11]

HV sublimation has been demonstrated to be an efficient method to grow very pure films of SCO nanomaterials with a sub-monolayer precision. This has been demonstrated for a reduced number of neutral mononuclear $Fe^{II}$ SCO complexes affording intermolecular interactions that favour sublimation in relatively mild conditions. Among them $[Fe(phen)_2(NCS)_2]$[12–14], $\{Fe[H_2B(pz)_2]_2(L)\}$[15–17] with L = phen and bipy ($[H_2B(pz)_2]^-$ = hydrobispyrazolylborate, L= 1,10-phenanthroline, 2,2'-bypiridine) and $[Fe(Pyrz)_2]$[18,19] (Pyrz = hydro-tris-(3,5-dimethyl-pyrazolyl)borate) have been most investigated. To the best of our knowledge, the first thin films based on SCO molecules deposited by sublimation were reported in the last decade for the archetypal $[Fe(phen)_2(NCS)_2]$[20]. More recently, this study was extended to investigate the effect of different metallic substrates on the SCO properties of this molecule.[21] However, thin films of the complex $\{Fe[H_2B(pz)_2]_2(phen)\}$ deposited on different substrates have been the most intensively investigated. These studies have revealed electron-induced reversible switching of single SCO complexes,[22] light induced excited spin state trapping (LIESST) phenomenon on ultrathin films deposited on Au(111),[23] also on highly oriented pyrolytic graphite (HOPG) surfaces,[24] and even electrically-sensed when embedded in a vertical junction.[25] In addition, they have inspired the preparation of homologous complexes derived from the 4-methyl-/chloro- and 4,7-dimethyl-/dichloro substituted phen,[26] whose SCO properties have been investigated in bulk and thin films by sublimation on HOPG and for the 3,4,7,8-tetramethyl homologous deposited on Au(111) and bismuth.[27] Concerning $\{Fe[H_2B(pz)_2]_2(bipy)\}$, it has been observed the locking and unlocking of the SCO behavior as a function of the film thickness on Au(111), or deposited on dielectric substrates ($SiO_2$ and $Al_2O_3$).[28–30] Also, a practically complete SCO transition has been observed on HOPG showing a clear growth of cooperativity with the increase of the number of monolayers from 0.35(4) to 10(1).[31] A different approach to explore cooperativity has been carried out preparing 10 nm thin films of the sublimed $\{Fe[H_2B(pz)_2]_2(bipy*)\}$ system on $SiO_2$, being bipy* a 2,2'-bipyridine functionalized with a dodecyl ($C_{12}$) alkyl chain to favour intermolecular interactions.[32] Lately, much endeavours have been dedicated in the sublimation of $[Fe(Pyrz)_2]$. At first, when processed as thin films (ca. 100 nm thick regime), this system was observed to present an incomplete thermal SCO transition, yet in a similar temperature range than the bulk material without clear loss of cooperativity.[33] Furthermore, both LIESST and soft X-ray induced excited spin state trapping (SOXIESST), known to occur in the bulk, were as-well found in the films. Subsequent work revealed the coexistence of two polymorphs within the sublimed films of this molecule, where each of them appears to present a different SCO behaviour.[34] Then, using the most stable polymorph a spin-state device was first produced incorporating this molecule.[35] More recently, further work was devoted in this direction by exploiting the more effective light-induced SCO properties of the less stable polymorph with its integration in a first-of-its-kind contactless graphene-based horizontal device.[36] In parallel, ultrathin films (submonolayers) of this molecule have been studied on different substrates showing to experience a variety of surface induced effects such as electric field induced HS/LS ordered superstructures, quenching of the HS state, reverse LIESST effect, single-molecule spin state manipulation and negative differential resistance.[37–43] All these studies have evidenced the vast and rich SCO phenomenology available for each of the different molecular systems aforementioned. Nonetheless, notice that other sublimable SCO molecules have been reported as-well such as {Fe[HB(pz)$_3$]$_2$},[44,45] {Fe[HB(trz)$_3$]$_2$},[46] [Fe(dpepd)(NCS)$_2$],[47] [Fe(qnal)$_2$]·xCH$_2$Cl$_2$,[48] and [Fe(pypyr(CF$_3$)$_2$)$_2$(phen)][49] (all the ligands corresponding to these molecules are depicted in **Scheme 1**).

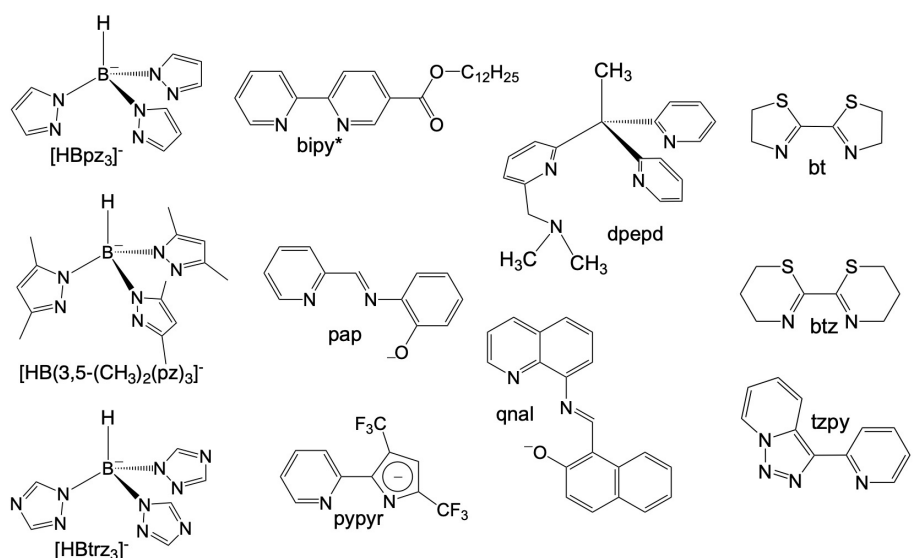

**Scheme 1**. Selection of ligands used in the preparation of sublimable Fe[II] complexes (see text).

In this context, here we describe the synthesis, structure, magnetic and photomagnetic properties for the bulk materials of three new SCO molecules of the {Fe[H$_2$B(pz)$_2$]$_2$(L)}·nS type, being L the bidentate-α-diimine ligands tzpy (**1·CH$_3$CN**, **1**), btz (**2**), and bt (**3·1/2CH$_2$Cl$_2$**, **3**) (see **Scheme 1**). Furthermore, we have succeeded in the preparation of thin films **TF1** and **TF2** deposited on different substrates (SiO$_2$, Au coated (40 nm) glass and CVD-graphene) from sublimation under HV conditions of as-synthesized **1** and **2** bulk powders respectively, probing their sublimable character. On the contrary, **3** has been found to decompose upon heating under HV conditions without subliming. The chemical integrity of **TF1** and **TF2** has been analyzed by means of infrared (IR) and Raman spectroscopies, while their thermal and light-induced SCO behavior has been monitored by means of the X-ray absorption spectroscopy (XAS). Finally, **TF1** and **TF2** films have been integrated on horizontal hybrid devices based on CVD-graphene to detect the thermal SCO behavior of the films through the electrical properties of the 2D material, since graphene has been previously proven as a highly sensitive platform to the changes undergone by SCO materials upon their spin transition (i.e. devices based on sublimed films and Van der Waals heterostructures).[35,36,50]

## RESULTS AND DISCUSSION

**Synthesis**.

The complexes {Fe[H$_2$B(pz)$_2$]$_2$(L)} (L = btz, tzpy and bt) were synthesized following the method previously described for L = phen and bipy.[15] In the reported conditions, {Fe[H$_2$B(pz)$_2$]$_2$(tzpy)} is rapidly formed giving a pink-orange microcrystalline precipitate (**1**). Single crystals of its acetonitrile solvate (**1·CH$_3$CN**) were isolated in 48 h from liquid-to-liquid layering of a MeOH solution of [Fe$^{II}$/2[H$_2$B(pz)$_2$]$^-$] (bottom) and an CH$_3$CN solution of tzpy (top) separated by a mixture MeOH:CH$_3$CN. The thermal analysis of the resulting needle-shaped crystals shows a sharp loss of weight in the interval 390-410 K consistent with a mol of CH$_3$CN per mol of complex (**Figure S1a**). Powder X-ray diffraction patterns show that the desolvated form corresponds to the pink-orange microcrystalline phase (**1**) (**Figure S1b**). Compound {Fe[H$_2$B(pz)$_2$]$_2$(btz)} (**2**) precipitates as a deep blue highly crystalline product of medium solubility, while single crystals for appropriate structure analysis were obtained within 24 h from evaporation of the resulting methanolic mother liquor under an Ar stream. The same synthetic procedure used for L = bt, gave single crystals of the unsolvated complex {Fe[H$_2$B(pz)$_2$]$_2$(bt)} (**3**). However, given the relative low solubility of bt in MeOH and in order to increase the yield of the complex, CH$_2$Cl$_2$ was used as alternative solvent where bt is more soluble, thereby affording the corresponding hemisolvate form (**3·1/2CH$_2$Cl$_2$**). The thermal analysis shows that a loss of weight, consistent with one molecule of CH$_2$Cl$_2$ per two moles of complex, takes place in the temperature interval 379-450 K (**Figure S2**).

**Magnetic properties**

The magnetic properties were monitored through the thermal dependence of the product $\chi_M T$ being $\chi_M$ the molar magnetic susceptibility and T the temperature. The $\chi_M T$ vs. T plots for the solvate and unsolvate form of the tzpy derivative are shown in **Figure 1a**. At 300 K, the $\chi_M T$ value is ca. 3.80 cm$^3$ K mol$^{-1}$ for **1** and **1·CH$_3$CN**. This value, decreases by ca. 0.3 cm$^3$ K mol$^{-1}$ in the upon cooling down to 180 K (**1**) and 200 K (**1·CH$_3$CN**). Below these temperatures both forms show a complete HS-to-LS transformation reaching a $\chi_M T$ value about 0.1 cm$^3$ K mol$^{-1}$ at 50 K. The SCO takes place, respectively, in one and two steps for the **1·CH$_3$CN** and **1**, being more cooperative for the latter. Indeed, in the heating mode the $\chi_M T$ vs T plot match quite well the cooling mode for the **1·CH$_3$CN** form but a very narrow hysteresis (ca. 4 K) is observed for **1**. The equilibrium temperature T$_{1/2}$ at which the HS and LS molar fractions are equal to 0.5 is 135 K (**1·CH$_3$CN**) and 167.5 K (**1**). For the latter, the characteristic temperatures at which the [$\partial(\chi_M T)/\partial T$] vs. T plot shows a maximum are 157 and 177K.

For compound **2**, the $\chi_M T$ value is 3.7 cm$^3$ K mol$^{-1}$ at 300 K and decreases slightly upon cooling down to 210 K (**Figure 1b**). Just below this temperature undergoes a cooperative one step SCO transition attaining a $\chi_M T \approx 0$ cm$^3$ K mol$^{-1}$ at 192 K. In the heating mode $\chi_M T$ vs. T plot does not match that of the cooling mode defining a hysteresis loop 6 K wide with (T$_{1/2}$)$^{down}$ = 197 K and (T$_{1/2}$)$^{up}$ = 203 K.

Concerning compound **3**, the $\chi_M T$ value 3.8 cm$^3$ K mol$^{-1}$ remains almost constant down to 50 K and then slightly decreases due to the occurrence of zero-field-splitting of the S = 2 HS state denoting the stabilization of the HS state at all temperatures. In contrast, for the solvate form **3·1/2CH$_2$Cl$_2$**, $\chi_M T$ decreases as T decreases in a very similar way as described for the solvate form of the tzpy, **1·CH$_3$CN**, undergoing a complete SCO characterized by T$_{1/2}$ = 116 K (**Figure 1c**)

Photo-generation of the metastable HS* state from the LS state, the so-called LIESST experiment,[51] was carried out at 10 K irradiating microcrystalline samples of the title compounds with green light (λ = 532 nm). In these conditions, the four samples that display SCO undergo LIESST effect with different yield. Indeed, $\chi_M T$ saturates to values of 3.4 cm³ K mol⁻¹ for **1** and **1·CH₃CN**, 1.9 cm³ K mol⁻¹ for **3·1/2CH₂Cl₂** and 0.9 cm³ K mol⁻¹ for **2**. Subsequently, the light was switched off and the temperature increased at a rate of 0.3 K min⁻¹ inducing a gradual increase of $\chi_M T$ that attains a maximum value of 3.8 cm³ K mol⁻¹ in the interval of 30-35 K for **1** and **1·CH₃CN**, 2.09 cm³ K mol⁻¹ for **3·1/2CH₂Cl₂** in the interval 20-30 K, and 0.97 for **2** at 17 K, which corresponds respectively to ca. 100%, 54% and 26% of the maximum value observed at 300 K. This raise in $\chi_M T$ reflects the thermal population of different microstates originated from the zero-field splitting of the HS* state. At higher temperatures, $\chi_M T$ decreases rapidly until joining the thermal SCO curve, respectively, at ca. 65, 63 and 47 K, indicating that the metastable HS* state has relaxed back to the stable LS state. The corresponding $T_{LIESST}$ temperatures, evaluated as $\partial(\chi_M T)/\partial T$,[52] are respectively ca. 55, 57 and 36 K. These temperatures are consistent with the inverse-energy-gap law, i.e. the metastability of the photo-generated HS* species decreases as the stability of the LS increases.[53–55]

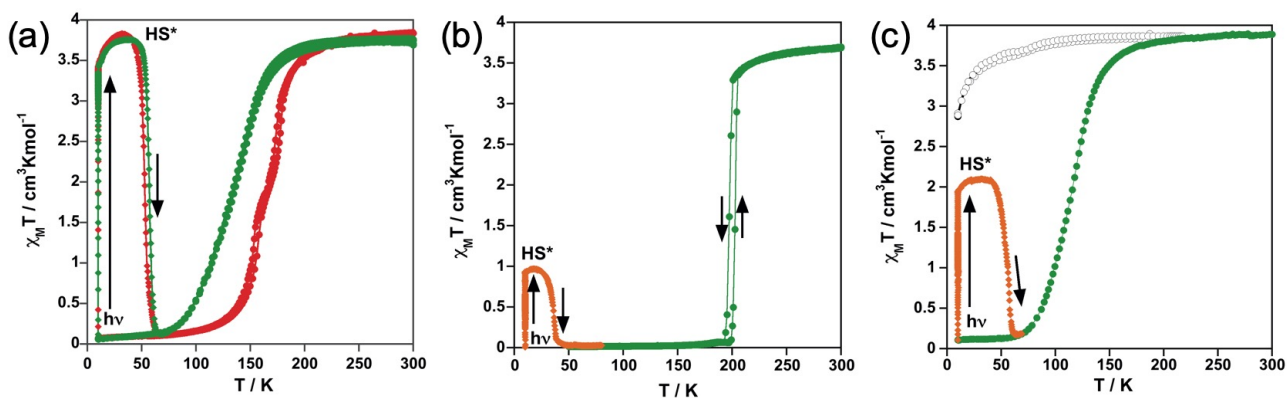

**Figure 1**. Magnetic and photomagnetic properties for (a) **1** and **1·CH₃CN** red and green solid circles, respectively; (b) **2** (green solid circles) and (c) **3·1/2CH₂Cl₂** (green solid circles) and **3** (black open circles).

### Structure

Single crystals of **1·CH₃CN**, **2**, **3·1/2CH₂Cl₂** and **3** were measured, respectively, at 110 K/250 K, 120 K/240 K, 120 K/220 K and 120 K. The crystals keep the same space group at both temperatures, **1·CH₃CN** (triclinic *P*-1), **2** (orthorhombic *Pna*2₁) and **3·1/2CH₂Cl₂** (monoclinic *P*2₁/*c*). The unsolvate form **3** adopts the triclinic *P*-1. Selected crystal data are gathered in **Table S1**.

The molecular structure for the four derivatives is depicted in **Figure 2** together with the atom numbering. The Fe$^{II}$ centre is surrounded by two [H₂B(pz)₂]⁻ anionic ligands adopting each other a *cis* conformation while the remaining positions are occupied by the bidentate-α-diimine ligand L = tzpy, btz or bt, thereby completing a distorted [FeN₆] octahedral geometry with the Fe-N bond lengths defined by the [H₂B(pz)₂]⁻ ligands shorter than those defined by L. A selection of significant bond-lengths and angles is gathered in **Tables S2-S6**. Furthermore, the complexes are chiral but both enantiomers coexist in the unit cell. Two slightly different Fe sites (Fe1 and Fe2) and consequently four crystallographically distinct [H₂B(pz)₂]⁻ ligands are found for **1·CH₃CN** (only site Fe1 site is displayed in **Figure 2a**, site Fe2 is shown in **Figure S3**). At 250 K, the average bond lengths <Fe1-N> = 2.188(3) Å and <Fe2-N> = 2.187(3) Å are practically identical. However, at 110 K both sites exhibit an unsymmetrical decrease of <Fe-N> as the LS state is populated: <Fe1-N> = 2.018(6) Å and <Fe2-N> = 2.058(6) Å, which involves a change in <Fe1-N> and <Fe2-N> of 0.170 and 0.129 Å, respectively. Taking into account that a structural reinvestigation of the L = phen and bipy homologous complexes shown that the <Fe1-N> change is 0.180 Å,[17] and assuming applicable this value to the title compounds, a transformation of 95% and 71% is found for Fe1 and Fe2 sites respectively. Considering the two Fe sites, the overall transformation is 83%, a value reasonably consistent with the $\chi_M T \approx 0.83$ cm³ K mol⁻¹ at 110 K, which corresponds to 78.2% of the HS-to-LS transformation. The angular distortion of the [FeN₆] from the ideal octahedron, defined as the sum of the angular deviation of the 12 *cis* angles ($\Sigma = \sum_{1}^{12}[\theta - 90]$) is very similar for the two sites $\Sigma_{Fe1}$ = 39.9° (110 K) and 58.2° (250 K) and $\Sigma_{Fe2}$ = 44.5° (110 K) and 52.3° (250 K).

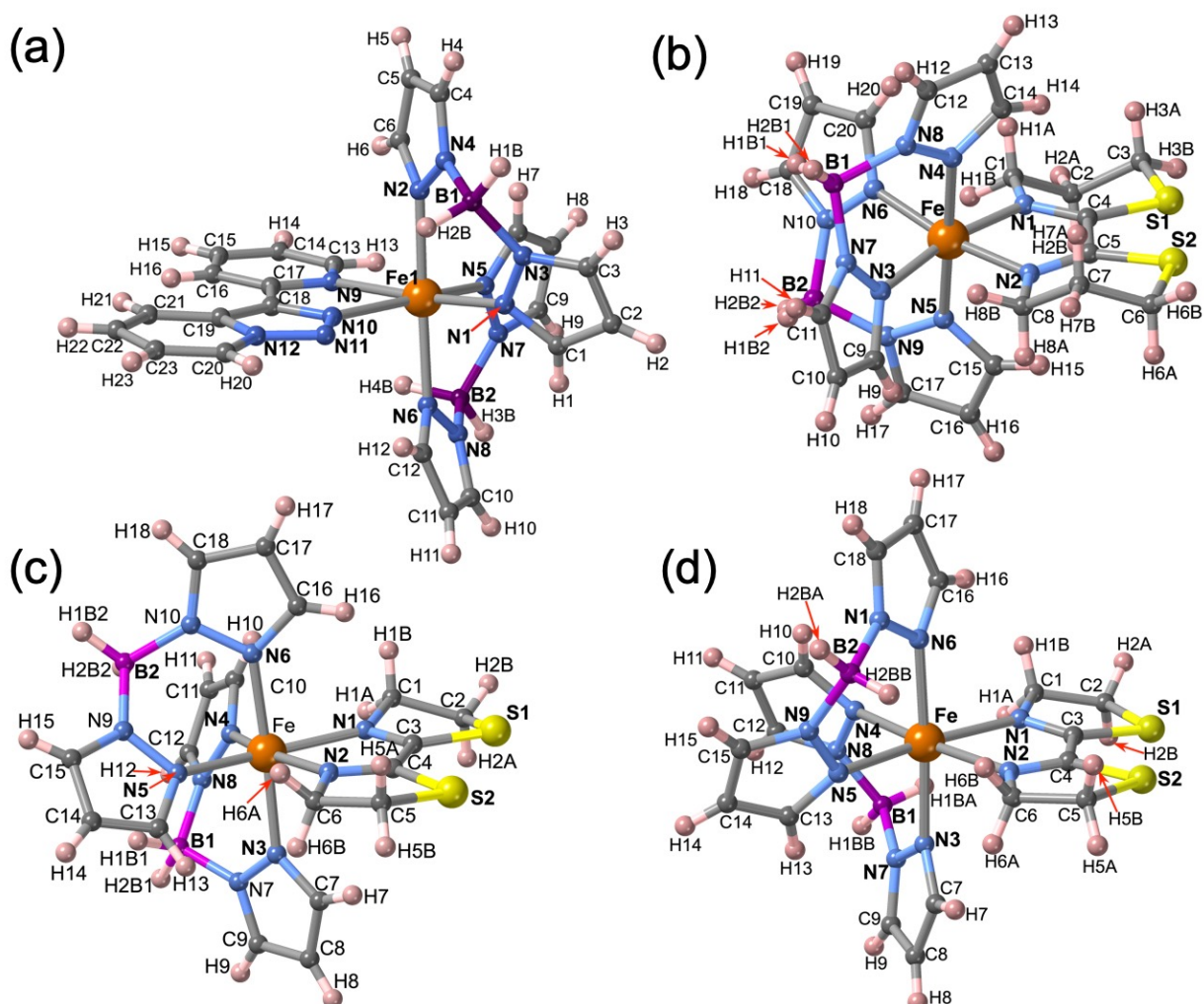

**Figure 2**. Molecular unit of **1·CH₃CN**, **2**, **3·1/2CH₂Cl₂** and **3** showing the atom numbering. Only one of the two Fe^II centres (Fe1) is shown in **1·CH₃CN**, for Fe2 see **Figure S3**.

The only one [FeN₆] octahedral site found for **2** is characterized by <Fe-N> = 1.985(4) and 2.172(5) Å at 120 and 240 K, respectively. The difference of 0.187 Å well-corresponds with the change of the average Fe-N bond-length typically observed for a complete HS↔LS transformation in agreement with the magnetic data. Furthermore, the SCO is accompanied by a change of $\Sigma_{Fe}$ from 47.3° (120 K) to 60.1° (240 K. For **3·1/2CH₂Cl₂** the <Fe-N> value is 2.181(4) Å at 220 K where the complex is fully

HS, and the decreases down to <Fe-N> = 2.103(4) Å at 120 K, the lowest attained temperature due to experimental limitations. The difference of 0.078 Å, consistent with a partial HS→LS transformation ca. 42%, matches reasonably well with the $\chi_M T$ value 2.3 cm³ K mol⁻¹ at 120 K, which corresponds ca. 40% of the complete HS→LS spin state change. In this interval of temperatures $\Sigma_{Fe}$ changes from 40.4° (120 K) to 52.1° (220). For the unsolvate form **3** the <Fe-N> value equal to 2.184(4) Å at 120 K is perfectly consistent with the HS shown by this compound at any temperature, in spite of the $\Sigma_{Fe}$ = 42.9°. However, at molecular level the most remarkable difference between the structure of **3** and its solvated homologue is the different relative conformation of the two [H₂B(pz)₂]⁻ ligands around the Fe$^{II}$ centre (see **Figure 3**). This difference makes the separation between the [H₂B(pz)₂]⁻ tetrahedrons significantly longer for **3** (6.585 Å) than for its solvate counterpart (4.607 Å). This "short" conformation is also observed for **2** while the "long" conformation is observed for **1·CH₃CN** as well as for the previously reported 2,2'-bipy and phen derivatives.

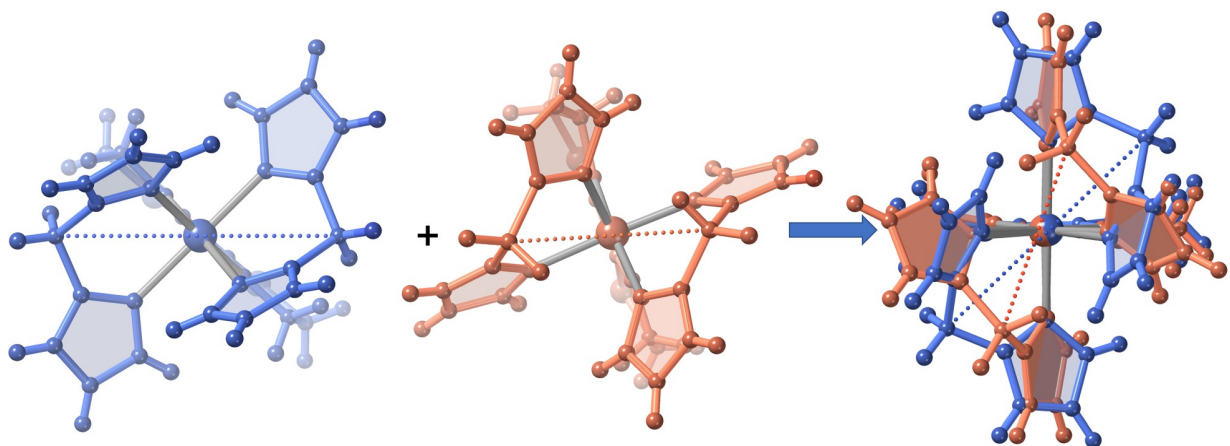

**Figure 3.** Superposition of the isomeric {Fe[H₂B(pz)₂]₂(bt)} complexes from **3** (blue color) and **3·1/2CH₂Cl₂** (salmon color) denoting the different conformations of the [H₂B(pz)₂]⁻ ligands around the Fe$^{II}$ centres (see text). Dotted lines visually show the distinct separation between the B tetrahedral centres of [H₂B(pz)₂]⁻.

The crystal packing of **1·CH₃CN** is depicted in **Figure 4a**. The unit cell contains four molecules, two for each crystallographically distinct Fe$^{II}$ centre. Similar to the {Fe[H₂B(pz)₂]₂(phen)}[15–17] complex, the aromatic nature of the tzpy ligand enables the formation of a very efficient π-π stacking, involving a great number of intermolecular C···C contacts smaller than the sum of the van der Waals radii (≈3.7 Å), between adjacent complexes with alternating Fe1-Fe2-Fe1 centres running along *a*-direction (**Figure 4b**). The resulting supramolecular zig-zag chains stack down *b*-direction, where no relevant interactions exist, defining layers parallel to the *a-b* plane which stack along c-direction generating voids where the CH₃CN molecules are located. Interestingly, the CH₃CN methyl group interacts with one of the pyrazole rings of each Fe1 and Fe2 centres (**Figure 4c**). The number and the length of these short contacts strongly depend on the spin state of the compound (see **Table S7**).

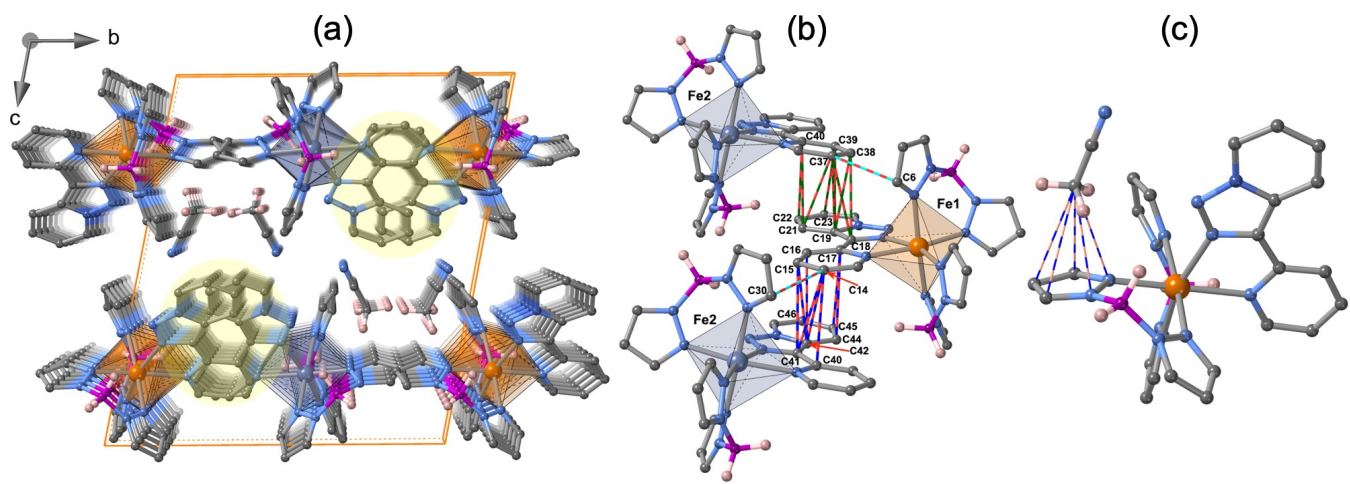

**Figure 4**. (a) Crystal packing of **1·CH₃CN** denoting the alternating Fe1 and Fe2 sites as orange and blue octahedrons. The strong π-π interactions are marked within a filled yellow circle. (b) Fragment of the supramolecular chains held together through strong π-π interactions. (c) Intermolecular interaction between the NCCH₃···pyrazole. Thin bicolor red-green/red-blue/blue-pink bonds represent the interatomic contacts smaller than the sum of the van der Waals radii.

The non-aromatic nature of the neutral α-diimine ligand btz is the cause of the much less compact molecular packing exhibited by **2**. Indeed, no relevant intermolecular contacts are observed for this compound even in the LS state at 120 K (**Figure S4** and **Table S8**). In contrast, the crystal packing of **3·1/2CH₂Cl₂** and **3** displays a larger number of intermolecular C···C and C···S contacts, particularly at 120 K, despite also being bt a non-aromatic ligand (**Figure S5** and **Table S8**).

**Sublimation of 1 and 2 as molecular thin films**

Thin films **TF1** and **TF2** ca. 200 nm thick were prepared by sublimation under HV conditions of the respective as-synthesized desolvated bulk powders **1** and **2** (see **Methods**). Also, sublimation of **3** was explored unsuccessfully since the bulk powder decomposed in the crucible upon heating without accomplishing any deposition onto overlaying substrates. Different substrates (SiO₂, Au coated (40 nm) glass and CVD-graphene) were used for the deposition of the molecular films depending on their purpose. It is important to note that in this thickness range the chemical integrity of SCO molecules is usually preserved for this family of compounds since surface-induced effects have only been observed at (sub)monolayer regime.[23,26,27,29,56–58] Furthermore, all films show similar morphology, regardless the type of substrate used, with highly homogenous coverages and very low roughness (ca. 1 nm), according to optical and atomic force microscopy (AFM) (**Figure 5**). These results match with the characteristics observed for other reported materials of this family of compounds in this thickness regime.[25,29,56,57]

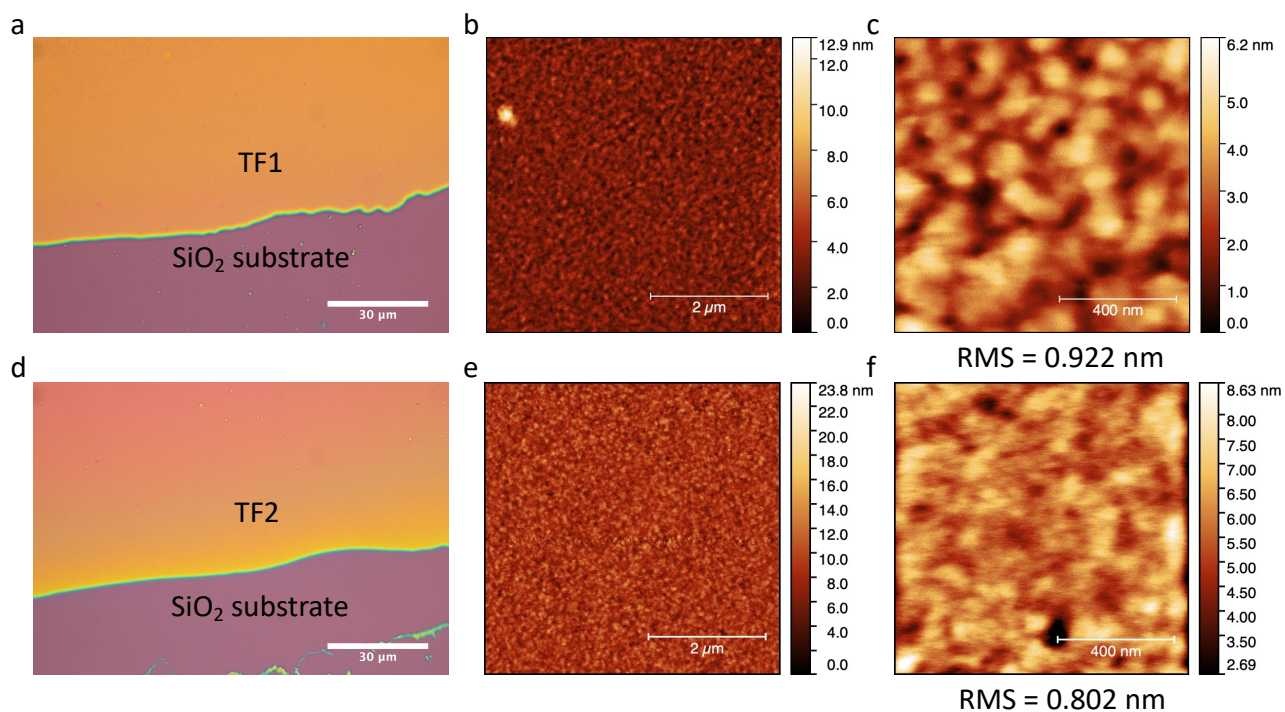

**Figure 5.** Optical microscope image (a), AFM image of 5 μm x 5 μm area (b) and AFM image of 1 μm x 1 μm area (c) collected on **TF1** deposited on $SiO_2$. Optical microscope image (d), AFM image of 5 μm x 5 μm area (e) and AFM image of 1 μm x 1 μm area (f) collected on **TF2** deposited on $SiO_2$. The calculated roughness of the films obtained as RMS value is indicated below each respective AFM image in c) and f). The scale in a) and d) is 30 μm, in b) and e) is 2 μm and in c) and f) is 400 nm.

The chemical integrity of **TF1**, **TF2** was analyzed by means of infrared (IR) and Raman spectroscopies. Moreover, their respective bulk materials (**1** and **2**) were characterized simultaneously for comparison. The high resemblance observed between the IR spectra of **TF1** and **TF2** grown on Au substrates, and those of their respective bulk references **1** and **2**, stand out as an indicative feature of the preservation of the chemical integrity of the molecules after their sublimation.[24] Note in particular the characteristic vibrational modes of the asymmetrical/symmetrical stretching of the B-H bond: 2407/2283 cm$^{-1}$ (**1**) and 2403/2283 cm$^{-1}$ (**2**) (**Figure 6a,b**). Similarly, the Raman spectra collected with a 532 nm laser for **TF1** and **TF2** grown on $SiO_2$ substrates using two different laser powers show high similarity to those of **1** and **2** (**Figure 6c,d**), evidencing retainment of the molecular integrity of the SCO molecules after sublimation, alike other systems of the same family of SCO molecules.[57]

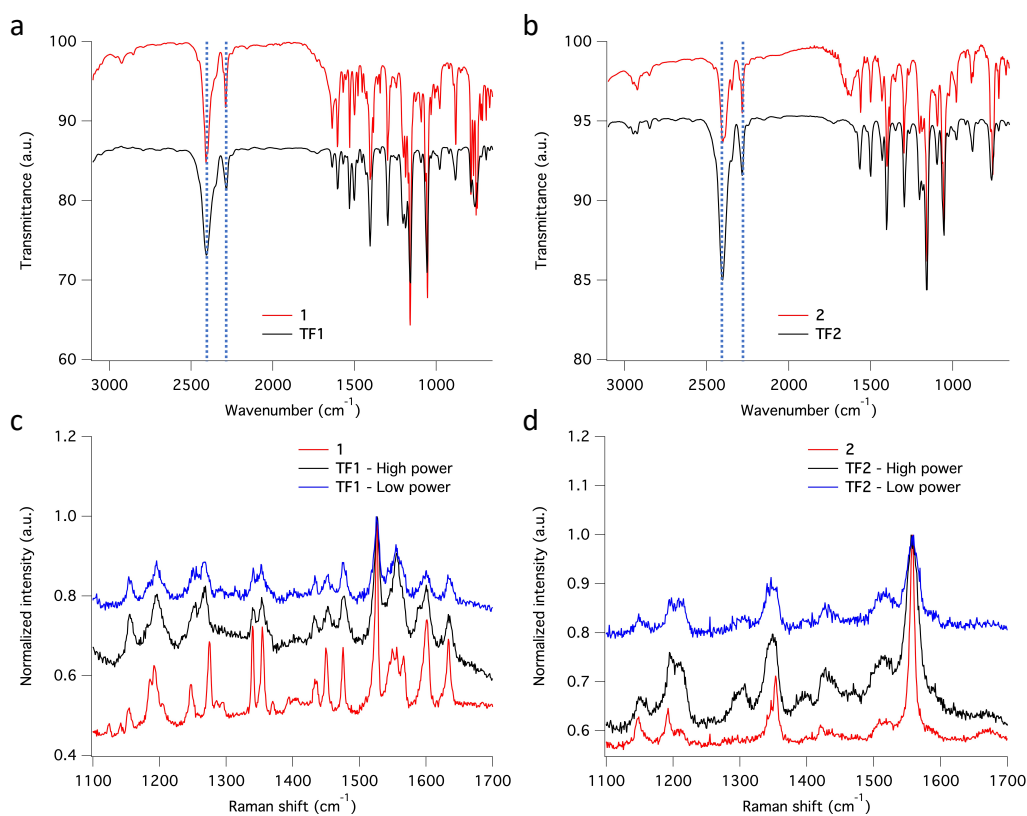

**Figure 6.** IR spectra collected for a) **1** embedded in a KBr pellet (red line) and **TF1** deposited on Au (black line) and b) **2** embedded in a KBr pellet (red line) and **TF2** deposited on Au (black line). Raman spectra collected for c) **1** scattered on a glass slide (red line – using a power of 20 μW·μm⁻²) and **TF1** deposited on SiO₂ (black line – using a laser power of 108 μW·μm⁻² and blue line - using a laser power of 11 μW·μm⁻²) and d) **2** scattered on a glass slide (red line – using a power of 20 μW·μm⁻²) and **TF2** deposited on SiO₂ (black line – using a laser power of 108 μW·μm⁻² and blue line - using a laser power of 11 μW·μm⁻²).

### Spin Crossover behavior of the deposited films TF1 and TF2 on SiO₂ and graphene substrates

The SCO behavior of the films **TF1** and **TF2** deposited on SiO₂ substrates, as well as that of their bulk counterparts **1** and **2**, was monitored through XAS technique at the Fe L$_{2,3}$ edges (see **Figure 7**). In order to get the thermal induced SCO profile of the films, temperature variable XAS spectra were recorded in both cooling and heating modes. Additional XAS spectra were recorded to characterize the LIESST effect irradiating the films for 10 min at 2 K with a red laser (λ = 633 nm) and the subsequent thermal relaxation of the photo-generated HS molecules upon heating in the dark. Concerning the bulk powders, only temperature variable XAS spectra were recorded during cooling mode.

Upon cooling, the almost disappearance of the peaks found at 708.1 and 708.9 eV of the Fe L$_3$ edge at high temperature (characteristic of HS state) in favor of the appearance of a very intense new peak at 709.6 eV at low temperature (characteristic of LS state) in **1** and **2** are consistent with the thermal spin transition (**Figure 7a,b,g,h**).[24,26,27,58,59] Furthermore, this is accompanied by a drastic increase in intensity of the main peak of the Fe L$_2$ edge at 721.5 eV. Therefore, the thermal dependence of the HS fraction for **1** and **2** was obtained for the cooling mode from a linear fitting of the XAS spectra of the fully populated HS and LS states, see **Figures 7c,i**. These thermal dependences characterized by $T_{1/2} \approx 170$ and 195 K, for **1** and **2**, respectively, are fully consistent with the cooling branches derived from the magnetic susceptibility measurements (**Figure 1a,b**). Nonetheless, the measured spectra also reveal partial oxidation of the bulk material since some contribution of $Fe^{III}$ is observed at the Fe $L_3$ edge at 710 eV invariably at all temperatures.[60] Further evidence is found at the Fe $L_2$ edge, where its shape corroborates this partial oxidation of the bulk material.[60] This observation is in strong contrast with the magnetic susceptibility measurements, since no oxidation is observed for neither compound as the low-temperature $\chi_M T$ values practically equal to zero are consistent with 100% of $Fe^{II}$ centres in both derivatives. Hence, we believe this partial oxidation is consequence from the unavoidable manipulation and adhesion of the bulk powders onto C-tape for the performance of the XAS experiment.

Regarding thin films, the recorded XAS spectra at 300 K show a pure HS-$Fe^{II}$ signal for both **TF1** and **TF2** films (red lines in **Figure 7d,e** and **Figure 7j,k** respectively). The lack of oxidation in this case is accomplished by keeping the films under inert atmosphere during the whole process, something unavoidable for the bulk powders. Upon cooling, both films appear to undergo some changes at the Fe $L_{2,3}$ edges in similar fashion to the above-described for the respective bulk powders. However, in this case, the total changes accomplished at low temperature (80 K and 50 K for **TF1** and **TF2** respectively) are much smaller (black lines in **Figure 7d,e** and **Figure 7j,k**). These results indicate that the thermal spin transition undergone by the films is incomplete, more gradual (blue lines in **Figure 7f**, **7l**) and shifted to lower temperatures, with $(T_{1/2})^{down} \approx 130$ and 140 K for **TF1** and **TF2**, respectively. Note that the exact value of the minimum HS fraction reached cannot be calculated with exact accuracy since the partial oxidation of the bulk powder prevented from collecting reliable pure HS and LS referential spectra. Nonetheless, from similar reported XAS characterization of molecules of this family we can estimate a maximum thermally achieved HS to LS conversion close to 20 % for **TF1** at 80 K and close to 25 % for **TF2** at 50 K.[24,26,27,58,59] Upon further cooling both molecules appear to experience some soft X-ray induced excited spin state trapping (SOXIESST) effect during the collection of the respective XAS spectra (blue lines in **Figure 7d,e** and **Figure 7j,k**), reaching at 2 K ca. 8 % of the initial HS state (blue lines in **Figure 7f** – between 2 and 80 K – and **Figure 7l** – between 2 and 50 K). This manifests the susceptibility of the films to soft X-rays at low temperatures (below 80 K for **TF1** and 50 K for **TF2**), something expectable since SOXIESST effect has been already observed for films of other molecules of this family.[26,27,58]

Furthermore, upon red laser irradiation at 2 K during 10 min, LIESST effect is accomplished (orange lines in **Figure 7d,e** and **Figure 7j,k**). This is especially effective in the film **TF2**, where ca. 70% of the LS molecules switch to the photo-generated HS state (orange line in **Figure 7f**), in contrast to the film **TF1** in which 40% of the LS centres are excited (orange line in **Figure 7l**). The metastable photo-generated HS states relaxed back to the LS through smooth heating (grey lines in **Figure 7d,e** and **Figure 7j,k**). This progression is clearly monitored between 2 K and 80 K (red lines in **Figure 7f** and **Figure 7l**) reaching back the ground state at ca. 80 K. Then, upon further heating, the thermal spin transition of both films shows full reversibility describing a similar trend as the cooling process, almost without hysteresis (red lines in **Figure 7f** and **Figure 7l**).

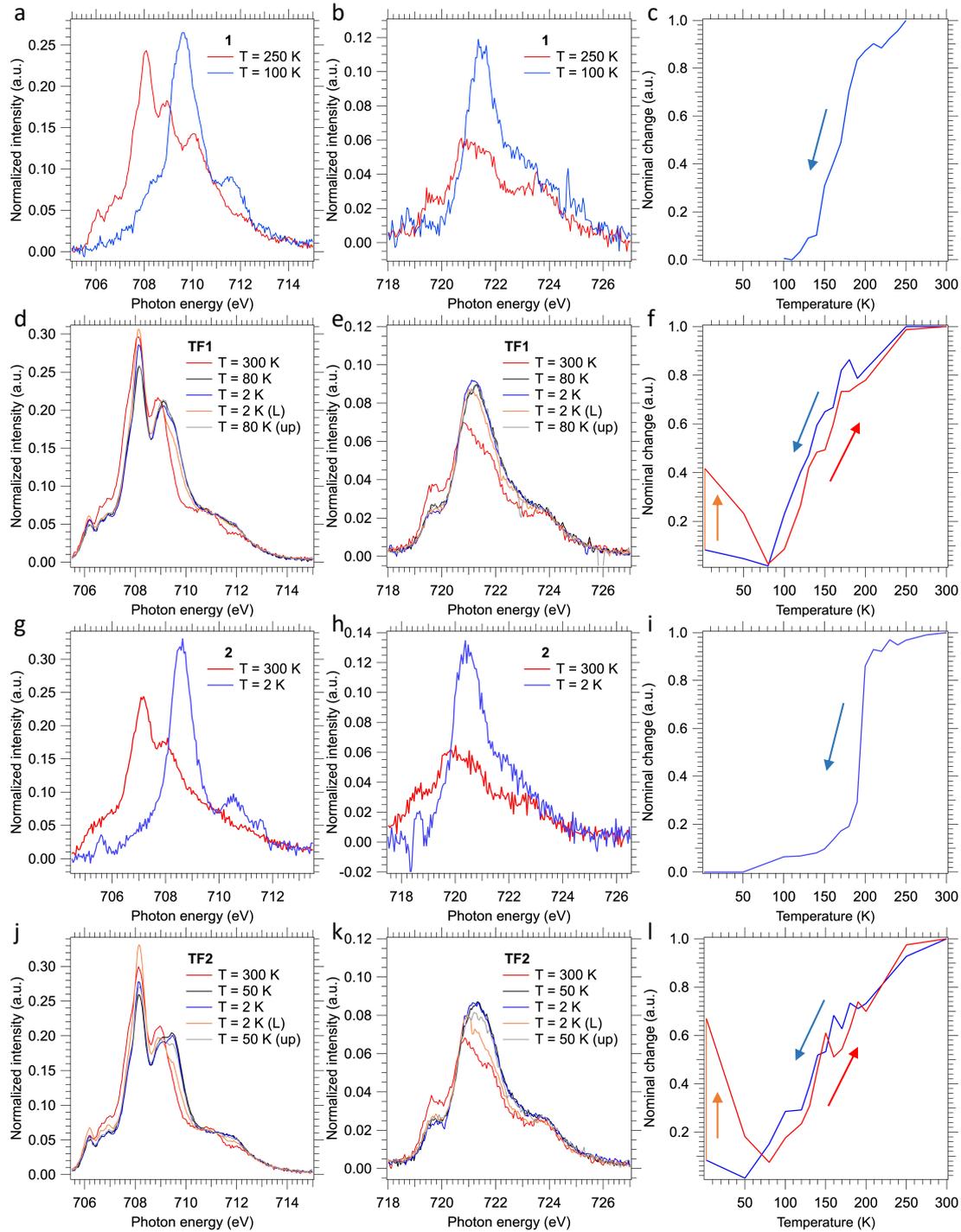

**Figure 7.** a) XAS spectra collected at 250 and 100 K in the Fe a) $L_3$ and b) $L_2$ edges energy range for **1** scattered on C-tape and c) calculated relative change from XAS spectra collected at each temperature for a cooling process in a temperature range varying from 250 to 100 K in 10 K steps. XAS spectra collected at 300, 80 and 2 K during a cooling process, 2 K after 10 min 633 nm laser irradiation, and 80 K during a heating process in the Fe d) $L_3$ and e) $L_2$ edges energy range for **TF1** deposited on $SiO_2$ and f) calculated relative change from XAS spectra collected at each temperature for cooling (blue line), 10 min 633 nm laser irradiation (orange line) and heating (red line) processes in a temperature range including 300, 250, and 220 K, a temperature ramp between 200 and 120 K in 10 K steps, and 100, 80, 50 and 2 K (before and after irradiation). XAS spectra collected at 300 and 2 K in the Fe g) $L_3$ and h) $L_2$ edges energy range for **2** scattered on C-tape and i) calculated relative change from XAS spectra collected at each temperature for a cooling process in a temperature range including 300, 275, and 250 K, a temperature ramp between 240 and 140 K in 10 K steps, and 120, 100, 50 and 2 K. XAS spectra collected at 300, 50 and 2 K during a cooling process, 2 K after 10 min 633 nm laser irradiation, and 50 K during a heating process in the Fe j) $L_3$ and k) $L_2$ edges energy range **TF2** deposited on $SiO_2$ and l)

calculated relative change from XAS spectra collected at each temperature for cooling (blue line), 10 min 633 nm laser irradiation (orange line) and heating (red line) processes in a temperature range including 300, 250, and 220 K, a temperature ramp between 200 and 120 K in 10 K steps, and 100, 80, 50 and 2 K (before and after irradiation).

SCO was also studied in hybrid interfaces formed by subliming **1** and **2** as thin films over graphene. Following a similar procedure to that recently reported by us,[36] horizontal devices were produced depositing **TF1** and **TF2** onto pre-contacted CVD-graphene. In these heterostructures, the resistive properties of the underlying CVD-graphene substrate are highly sensitive to the strain induced by the volume change associated with the spin state change of the anchored SCO layer.[35,36] The measurements of the transport properties of the devices as a function of temperature for a single full thermal cycle between 2 and 250 K are presented in **Figure 8**. In both cases a change in the resistance of graphene is observed in the SCO/graphene device that does not appear in pristine CVD-graphene devices.[36] Thus, upon cooling a progressive increase in the resistance is observed in the range 50-160 K centered at ca. 100-120 K, in good agreement with XAS measurements. Upon heating, an apparent hysteretic behavior is observed. However, this behavior is not intrinsic to the material but due to the progressive loss of mass in the film under HV conditions, as can be seen by its overall thinning after the thermal cycling by ca. 25-30% (**Figure S6**). In fact, the cycling involves long measurement times (for example, a full cycle at 1 K·min⁻¹ recording data every 0.5 K requires ca. 10 h), in such a manner that under HV conditions part of the film sublimes back. Nonetheless, these results account for the successful electrical detection of the spin crossover transition in these films, corroborating the XAS experiment.

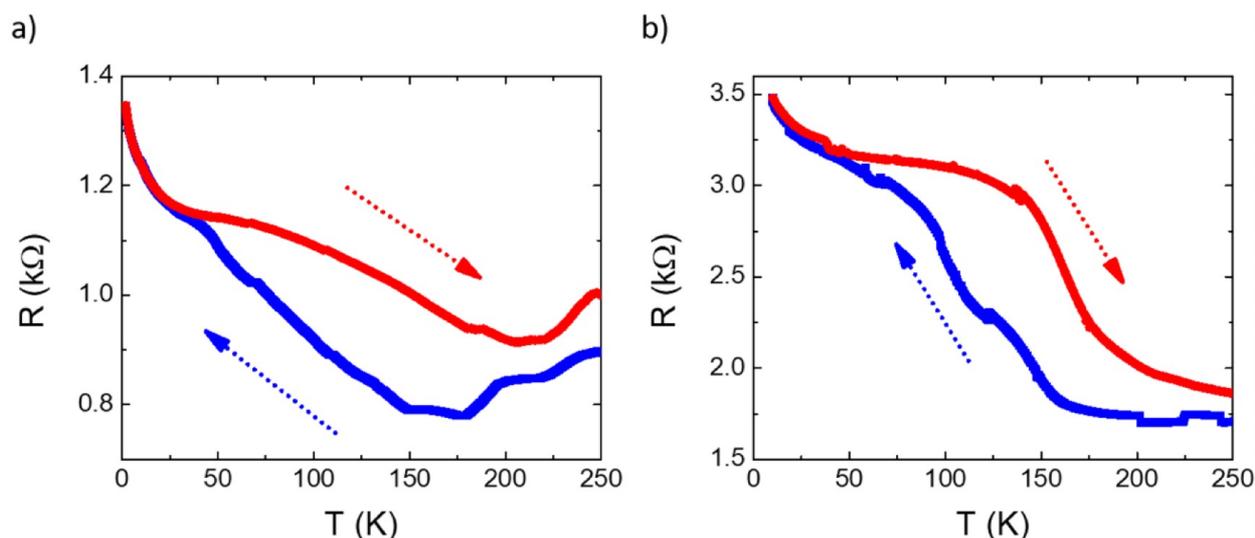

**Figure 8.** Temperature dependence of the resistance in in-contact CVD-graphene horizontal devices integrating a 200 nm thick films of a) **1** and b) **2** in the range between 2 and 250 K showing in blue the cooling process and in red the heating process.

## CONCLUSIONS

Sublimable spin crossover molecules are still very scarce. Here we have prepared novel molecules of this kind by extending the family {Fe[H$_2$B(pz)$_2$]$_2$(L)}, where L is a bidentate-α-diimine ligand. In a second step we assess their suitability to be integrated as thin films into resistive devices. The three chosen bidentate $\alpha$-diimine ligands, L = tzpy, btz and bt, have afforded the targeted complexes (**1-3**) plus two additional solvated forms (**1·CH₃CN**, **3·CH₂Cl₂**). All derivatives except **3·CH₂Cl₂** undergo thermal- and light-induced SCO behaviour with $T_{1/2}$ values in the range 115-200 K.

Sublimation of the two unsolvated SCO derivatives undergoing SCO properties, **1** and **2**, and subsequent controlled condensation of the molecules on specific substrates has afforded the corresponding 200 nm thick thin films **TF1** and **TF2** whose SCO properties have been characterized through XAS studies. From these XAS results, different features are observed in the SCO behavior that contrast with the expected as compared to the respective bulk powders: (i) the incomplete character of the thermal spin transition; (ii) the total loss of cooperativity; and iii) the shifting to lower temperatures. Several circumstances may converge in the origin of these features. First, the high sensitivity of the films to X-ray irradiation even at high temperatures (> 100 K). This effect could prevent a large fraction of the molecules from properly undergoing the thermal spin transition, as it has been previously observed in some bulk SCO materials and more recently for films of a SCO molecule, where the thermal spin transition was totally supressed.[36,61] Thus, since both films **TF1** and **TF2** show the typical SOXIESST effect up to 80 K and having a rather strong shifting to lower temperatures of their thermal spin transition, it could be reasonable to consider this high sensitivity to X-ray irradiation as possible cause. A second additional explanation may arise from the very small size of nanoparticles in the film, not distinguishable in the AFM images collected (**Figure 1b,c,e,f**), and the presence of defects, pinning some of these molecules in the HS state regardless the temperature.[56–58,62] A partial decomposition of the pristine molecules on the surface to form tetrahedral (SCO inactive) [Fe(bpz)₂] fragments could be also considered.[27,58] However, since this fragmentation is only expected to occur at (sub)monolayer level and would also affect substantially to the shape of the XAS spectra at the Fe $L_{2,3}$ edge, this can be discarded in our case.

Finally, we have integrated these films **TF1** and **TF2** onto CVD-graphene based horizontal hybrid devices and detected their thermal SCO transition through the changes observed in the electrical properties of the underlying 2D material.

## ACKNOWLEDGEMENTS

This work was supported by: the European Union (ERC AdG Mol-2D 788222), the Spanish MICINN (2D-HETEROS PID2020-117152RB-100 and SUPERSCO PID2020-117264GB-100, co-financed by FEDER, Grant PID2019-106147GB-I00 funded by MCIN/AEI/10.13039/501100011033 and Excellence Unit "María de Maeztu", CEX2019-000919-M) and the Generalitat Valenciana (Prometeo program and PO FEDER Program, ref IDIFEDER/2018/061 and IDIFEDER/2020/063). M.G.-E. acknowledges the support of a fellowship FPU15/01474 from MIU. F.J.V.-M. acknowledges the support of the Generalitat Valenciana (APOSTD/2021/359). R.C. acknowledges the support of a fellowship from "la Caixa" Foundation (ID 100010434), LCF/BQ/PR19/11700008 and the Generalitat Valenciana (SEJIGENT/2021/012T). All XAS experiments were performed at Boreas beamline at ALBA Synchrotron with J.H.-M. in both proposal and in-house experiments. The authors thank Alejandra Soriano Portillo and Ángel López Muñoz for their technical support.

## METHODS

***Synthesis***

The ligands tzpy, btz and bt were synthesized according to methods previously described.[63,64] The potassium salt of the hydrobispyrazolylborate, K[H$_2$B(pz)$_2$], was purchased from commercial sources.

***Synthesis of {Fe[H$_2$B(pz)$_2$]$_2$(L)}***. All operations were performed in an Ar atmosphere. To a methanolic solution of K[H$_2$B(pz)$_2$] (4 mL, 98 mg, 0.526 nmol) prepared in a small Schlenk flask was added a colourless methanolic solution of Fe(ClO$_4$)$_2$·xH$_2$O (x ≈ 6) (2 mL, 100 mg., 0.263 mmol) containing some crystals of ascorbic acid dissolved. After stirring for 10 min a white-greyish powder appears (KClO$_4$) which was filtered off using a cannula under anaerobic conditions. Then, once transferred the resulting colourless solution 1Fe$^{II}$:2[H$_2$B(pz)$_2$]$^-$ in a larger Schlenk flask, a methanolic solution (4 mL) containing 0.263 mmol of L (tzpy, 51.4 mg; btz, 52.7 mg; bt, 45.24 mg) was added under continuous stirring. The tzpy derivative precipitated rapidly affording a pink-yellow microcrystalline powder which corresponds to the unsolvated form **1**. In contrast, single crystals of the unsolvated forms of the btz (**2**) and bt (**3**) derivatives were obtained from evaporation of the resulting dark-blue mother liquor under argon flow in 24 h. Anal. Calcd for C$_{23}$H$_{24}$N$_{12}$B$_2$Fe (**1**): C, 50.60; H, 4.43; N, 30.78. Found: C, 51.26; H, 4.48; N, 31.18. Anal. Calcd for C$_{20}$H$_{28}$N$_{10}$B$_2$S$_2$Fe (**2**): C, 43.67; H, 5.13; N, 25.46. Found: C, 44.07; H, 5.20; N 25.73. Anal. Calcd for C$_{18}$H$_{24}$N$_{10}$B$_2$S$_2$Fe: C, 41.41; H, 4.63; N, 26.83. Found: C, 41.65; H 4.52, ; N 27.12.

Single crystals of **1·CH$_3$CN** were obtained by slow liquid diffusion (layering method) using test tubes containing a MeOH solution (3mL) of 1Fe$^{II}$/2[H$_2$Bpz$_2$]$^-$ at the bottom, a 1MeOH:1CH$_3$CN mixture (10 mL) in the middle and a CH$_3$CN solution (2 mL) of tzpy on top. Single crystals of **3·1/2CH$_2$Cl$_2$** were accidentally obtained trying to improve the yield of **3**. The synthetic protocol was modified replacing the MeOH with CH$_2$Cl$_2$ as solvent and keeping the resulting mother liquor in an ice bath during its slow evaporation in argon stream. The crystals of **1·CH$_3$CN**, **2**, **3** and **3·1/2CH$_2$Cl$_2$** were analyzed by single crystal X-ray diffraction methods. The solvated compounds were also confirmed through thermogravimetric analysis.

***Thin film sublimation***: Thin films **TF1** and **TF2** were prepared by sublimation under high vacuum (HV) conditions of the as-synthesized and desolvated bulk powder of each of the molecules (**1** and **2** respectively, ca. 100 mg). For the film processing, these precursors were respectively introduced in quartz crucibles inside Knudsen cells fitting in our customized evaporation chamber (CREATEC Molecular Beam Epitaxy (MBE) customized system in a Clean Room Class 10000). The sublimation conditions consisted in: a) **1** heated at 100 °C in a 5·10$^{-7}$ mbar HV to reach a deposition rate of 0.4 Å/s; and b) **2** heated at 90 °C in a 5·10$^{-6}$ mbar HV to reach a deposition rate of 0.4 Å/s. These deposition rates were monitored in-situ sublimation through a calibrated quartz crystal microbalance (QCM) and the final thicknesses of the produced films (200 nm thick) were verified by profilometry (KLA Alpha-Step D-500 profilometer with nanometric resolution).

***Physical characterization***

*Elemental analyses* (C, H, and N) were performed with a CE Instruments EA 1110 CHNS Elemental analyzer. *Magnetic measurements* were performed with a Quantum Design MPMS-XL-5 SQUID magnetometer in the 2 to 400 K temperature range with an applied magnetic field of 1 T. Experimental magnetic susceptibilities were corrected for diamagnetism of the constituent atoms by the use of Pascal's constants. *Photomagnetic measurements* were performed irradiating with a Diode Pumped Solid State Laser DPSS-532-20 from Chylas and a coupled via an optical fiber to the cavity of the SQUID magnetometer. The optical power at the sample surface was adjusted to ~3 mW·cm$^{-2}$, and it was verified that it resulted in no significant change in magnetic response due to heating of the sample. The samples consisted of a thin layer of compound whose weight was corrected by comparison of a thermal spin crossover curve with that of a more accurately weighted sample of the same compound. *Calorimetric measurements* were performed using a differential scanning calorimeter Mettler Toledo DSC 821e. Low temperatures were obtained with an aluminium block attached to the sample holder, refrigerated with a flow of liquid nitrogen and stabilized at a temperature of 110 K. The sample holder was kept in a dry box under a flow of dry nitrogen gas to avoid water condensation. The measurements were carried out using around 10-15 mg of polycrystalline samples sealed in aluminium pans with a mechanical crimp. Temperature and heat flow calibrations were made with standard samples of indium by using its melting transition (429.6 K, 28.45 J g$^{-1}$). An overall accuracy of ±0.2 K in temperature and ±2% in the heat capacity is estimated. The uncertainty increases for the determination of the anomalous enthalpy and entropy due to the subtraction of an unknown baseline. *Powder X-ray measurements were* performed on a PANalytical Empyrean X-ray powder diffractometer (monochromatic CuKα radiation).

*Single crystal X-ray measurements:* Single crystals were mounted on a glass fiber using a viscous hydrocarbon oil to coat the crystal and then transferred directly to the cold nitrogen stream for data collection. X-ray data were collected on an Oxford Diffraction Supernova diffractometer equipped with a graphite-monochromated Enhance (Mo) X-ray Source (λ = 0.71073 Å). The program CrysAlisPro, Oxford Diffraction Ltd., was used for unit cell determinations and data reduction. Empirical absorption correction was performed using spherical harmonics, implemented in the SCALE3 ABSPACK scaling algorithm. The structures were solved by direct methods and refined by full matrix least-squares on $F^2$ using SHELXL-2018.[65] Non-hydrogen atoms were refined anisotropically, and hydrogen atoms were placed in calculated positions refined using idealized geometries (riding model) and assigned fixed isotropic displacement parameters. The CCDC files 2209227 (**3·1/2CH$_2$Cl$_2$**, 120 K), 2209228 (**2**, 120 K), 2209229 (**3**, 120 K), 2209230 (**2**, 240 K), 2209231 (**3·1/2CH$_2$Cl$_2$**, 220 K), 2209232 (**1·CH$_3$CN**, 110 K) and 2209233 (**1·CH$_3$CN**, 250 K) contain the supplementary crystallographic data for this article. These data can be obtained free of charge from The Cambridge Crystallographic Data Centre via www.ccdc.cam.ac.uk/data_request/cif.

*Thin film characterization*: Optical microscopy images were collected using a Nikon Eclipse LV-150N microscope equipped with a Nikon DS-FI3 camera through a 100x objective. *Atomic Force Microscopy* (AFM) images collected for **TF1** and **TF2** were performed using a Bruker Dimension Icon with Scan Assyst in tapping mode in 1 μm × 1 μm and 5 μm × 5 μm areas. These images were processed using Gwyddion software, from which the roughness of the films was obtained as Root Mean Square statistical value (RMS). *Infrared Spectroscopy*: IR spectra were collected using a Fourier

Transformation-Infrared Spectrometer NICOLET 5700 from Thermo Electron Corporation in the wavenumber range between 3100 cm$^{-1}$ and 650 cm$^{-1}$. For **TF1** and **TF2**, grown on 3 cm × 3 cm Au coated glass substrates, a module was used that allowed the measurement of the transmittance of the reflected IR light from the sample. In the case of the respective bulk powders, embedded in KBr pellets, a second module was used that allowed the measurement of the transmittance of the IR light after going through the KBr pellets separately containing **1** and **2** dispersed molecular materials. *Raman Spectroscopy*: Raman spectra were collected on **TF1** and **TF2** grown on 7 mm × 7 mm SiO$_2$ substrates and on **1** and **2** scattered powders on a glass slide using a Horiba LabRAM HR Evolution equipped with a 532 nm Laser beam with a maximum power of 1.08 mW·μm$^{-2}$. The Raman shift in these spectra ranged between 1100 cm$^{-1}$ and 1700 cm$^{-1}$. For the bulk powders the power used during the data collection was of 20 μW·μm$^{-2}$. In the case of the films, spectra with two different laser powers were collected: a) 11 μW·μm$^{-2}$ and b) 109 μW·μm$^{-2}$.

*X-ray Absorption Spectroscopy*: XAS characterization of **TF1** and **TF2** deposited on 7 mm × 3 mm Si/SiO$_2$ substrates and of **1** and **2** bulk powders was performed at Boreas beamline in ALBA synchrotron. The films were attached to copper sample holders using aluminum clips and placing indium foil below to improve their thermalization. Bulk powders were scattered onto C-tape placed directly on top of copper sample holders. The measurements implied the collection of several energy scans focused at the Fe L$_{2,3}$ edge region at different temperatures onto each sample, cooling and heating processes for the films and only cooling for the powders. These scans were performed using total electron yield mode and substantially low photon flux (intensity ≤ 0.025 nA) with a sufficient signal to noise ratio. Moreover, all samples were only irradiated with X-rays during the spectra collection, only exposing them during these periods to prevent as much as possible unwilled/avoidable X-ray induced phenomena. For the LIESST effect study on the films, these were irradiated with a red laser (He-Ne LASER from Research Electro-optics Inc. (R-30993), wavelength 633 nm and power 12 mW) at 2 K, right after the cooling process, for 10 min. All collected spectra for this study were processed for their analysis through background subtraction and normalization. The evolution of the spectra collected for each case with temperature was calculated respectively as relative change from its fitting to a linear combination of the spectra showing the highest (1 in the plots) and lowest (0 in the plots) HS fraction respectively.

*Devices fabrication and electrical characterization*: Both devices fabrication and electrical characterization were performed following the methods described in our previous work for in-contact devices but using the SCO molecules described in the present text with their respective sublimation conditions.[36]